\documentclass[preprint,showpacs,amsmath,amssymb]{revtex4}
\usepackage{mathrsfs}
\usepackage{graphicx}
\usepackage{amsfonts}
\usepackage{dcolumn}
\usepackage{bm}

\usepackage{CJK}

\begin{document}
\title{Epidemic extinction in a generalized susceptible-infected-susceptible model}

\author{Hanshuang Chen$^{1}$}\email{chenhshf@ahu.edu.cn}

\author{Feng Huang$^{2}$}

\author{Haifeng Zhang$^3$}\email{haifengzhang1978@gmail.com}

\author{Guofeng Li$^1$}

\affiliation{$^{1}$School of Physics and Materials Science, Anhui
University, Hefei, 230601, China \\ $^{2}$School of Mathematics and
Physics, Anhui Jianzhu University, Hefei 230601, China \\$^3$School
of Mathematical Science, Anhui University, Hefei, 230601, China}

\date{\today}

\begin{abstract}
We study the extinction of epidemics in a generalized
susceptible-infected-susceptible model, where a susceptible
individual becomes infected with the rate $\lambda$ when contacting
$m$ infective individual(s) simultaneously, and an infected
individual spontaneously recovers with the rate $\mu$. By employing
the Wentzel-Kramers-Brillouin approximation for the master equation,
the problem is reduced to finding the zero-energy trajectories in an
effective Hamiltonian system, and the mean extinction time $\langle
T\rangle$ depends exponentially on the associated action $\mathcal
{S}$ and the size of the population $N$, $\langle T\rangle \sim
\exp(N\mathcal {S})$. Because of qualitatively different bifurcation
features for $m=1$ and $m\geq2$, we derive independently the
expressions of $\mathcal {S}$ as a function of the rescaled
infection rate $\lambda/\mu$. For the weak infection, $\mathcal {S}$
scales to the distance to the bifurcation with an exponent $2$ for
$m=1$ and $3/2$ for $m\geq2$. Finally, a rare-event simulation
method is used to validate the theory.
\end{abstract}
\pacs{} \maketitle

\section{Introduction}
The extinction of epidemics in finite populations is one of major
challenges in population dynamics \cite{Anderson1991,Nature460.334}.
Although many factors may contribute, such as environmental changes
and social factors, intrinsic fluctuations, originated from
discreteness of the reacting agents and the random character of
their interactions, can induce a rare but large fluctuation along a
most probable, or optimal, path to extinction of epidemics
\cite{PhysRevLett.101.078101,PhysRevLett.103.068101,PhysRevLett.101.268103}.
Unlike the fluctuation probability in equilibrium systems determined
by the Boltzmann distribution, population dynamics is far away from
equilibrium, and therefore, a general principle to determine the
probability of large fluctuations in out-of-equilibrium system is
still lacking.

Many efforts have been done on this topic. Among them, many
theoretical works have been focused on the calculation of mean
extinct time (MET) of mathematical epidemic models, in which the
systems possess a stable size of epidemic population which is
metastable as rare large fluctuations always drive it to an
absorbing state without infective individuals in finite populations.
When the populations size is large, the Fokker-Planck approximation
to the exact master equation that describes the stochastic process
of the spreading only describes small deviations from the
probability distribution maxima, but it fails to determine the
probability of large fluctuations \cite{MultiscaleModelSimul3.283}.
Elgart and Kamenev made an important progress in this topic,
pioneered in
Refs\cite{JSP9.51,PhysRevA.36.5782,Bull.Math.Biol.51.625,JCP100.5735}:
they employed the Peliti-Doi technique
\cite{JPA9.1465,JPhysFrance46.1469,PhysRevLett.77.4780,JSP90.1}, to
map the master equation into a Schr\"odinger-like equation that can
identify the classical trajectory that connects the metastable fixed
point and the absorbing state \cite{PhysRevE.70.041106}. This allows
them to calculate the classical action along this trajectory, a
first approximation to the MET. Assaf and Meerson then suggested a
general spectral method to improve the Elgart-Kamenev results
\cite{PhysRevLett.97.200602,PhysRevE.75.031122}. Kessler and Shnerb
presented a general method to deal with the extinction problem based
on the time-independent ``real space" Wentzel-Kramers-Brillouin
(WKB) approximation \cite{JSP127.861}. This method is easy to
implement, its intuitive meaning is transparent, and its range of
applicability covers any single species problem. These novel methods
have been successfully applied to solve extinction problems in
diverse situations, including time-varying environment
\cite{PhysRevE.78.041123,PhysRevE.81.031126}, catastrophic events
\cite{PhysRevE.79.011127}, fragmented populations with migration
\cite{PhysRevLett.109.138104,PhysRevLett.109.248102}, complex
networks \cite{EPL108.58008,PhysRevLett.117.028302}, and some others
\cite{PhysRevE.77.061107,PhysRevE.81.051925,PhysRevE.83.011129,PhysRevE.85.021140,PhysRevE.87.032127,PhysRevE.93.032109}.

In this paper, we shall apply the WKB approximation to calculate the
MET in a variant of susceptible-infected-susceptible (SIS) model, in
which a susceptible individual becomes infected with the rate
$\lambda$ when contacting $m$ infective individual(s)
simultaneously, and an infected individual spontaneously recovers
with the rate $\mu$. The population is well-mixed. For $m=1$, the
model recovers to the standard SIS model. $m\geq2$ implies that a
susceptible individual can be infected only if (s)he contacts at
least two infective individuals. Since there are qualitatively
different bifurcation features between $m=1$ and $m\geq2$, we
calculate independently the MET, which is exponentially dependent on
the action $\mathcal {S}$ along the optimal extinction path and
population size $N$. Near the infection threshold, we find that
$\mathcal {S}$ scales to the distance to the threshold with an
exponent $2$ for $m=1$ and $3/2$ for $m\geq2$. Lastly, we will use a
rare-event simulation to validate the theoretical results.

\section{Model}
We consider a variant of the well-known SIS model. The model
consists of $N$ well-mixed individuals, in which each individual is
either susceptible ($S$) or infected ($I$). The dynamics of the
model is governed by the following reactions,

\begin{eqnarray}
S + mI &\mathop { \to }\limits^\lambda&  (m + 1)I \nonumber \\
I&\mathop  \to \limits^\mu &  S \nonumber
\end{eqnarray}
The first reaction describes that an individual $S$ is infected when
contacting at least $m$ individuals $I$ simultaneously with the
infection rate $\lambda$, and the second one refers to the recovery
process of an individual $I$ with the recovery rate $\mu$. Without
loss of generality, we set $\mu=1$. For a given integer $m$ no less
than one, the model is named as SIS$m$ for convenience. For $m=1$,
the model recovers to the standard SIS model. For $m=2$, the model
has been studied on two-dimensional lattices
\cite{PhysRevLett.98.050601,JSP135.77} and on complex networks
\cite{PhysRevE.87.062819,NJP17.023039}, which have found a
discontinuous phase transition can emerge. A common feature for any
$m$ is that there always exists an absorbing state corresponding to
not any more individuals $I$. This implies that the state cannot be
left once the dynamics brings the system into it.
\section{Mean-field theory}
We proceed our analysis from the deterministic mean-field theory.
The rate equation for the density $x$ of individuals $I$ can be
written as
\begin{eqnarray}
\dot x = \lambda (1 - x){x^m} -  x \label{eq1}
\end{eqnarray}
where $x=n/N$ and $n$ is the number of individuals $I$. It is
clearly noticed that $x=0$ is always a steady solution of
Eq.\ref{eq1}, corresponding to an epidemic-free state (absorbing
state).

\begin{figure}
\centerline{\includegraphics*[width=1.0\columnwidth]{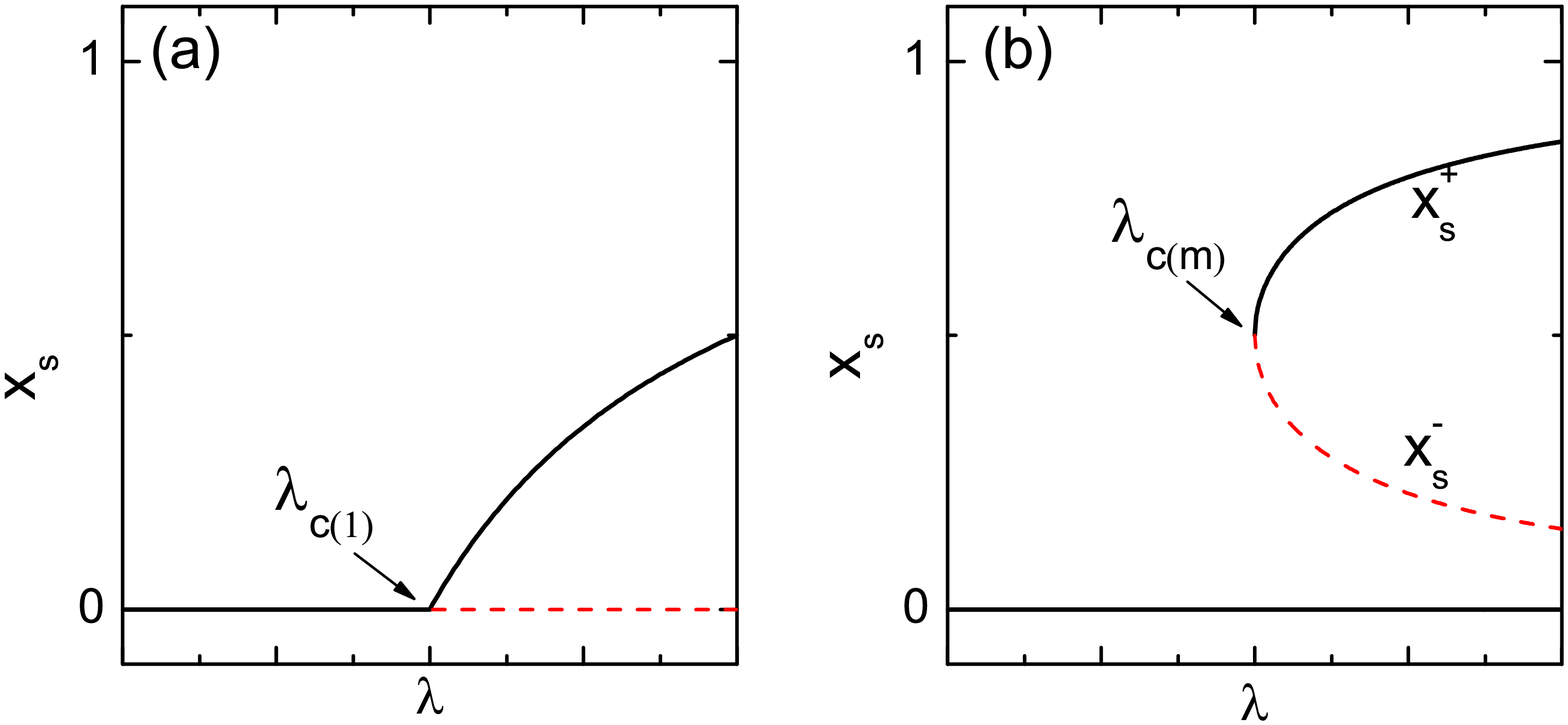}}
\caption{(color online). Bifurcation diagrams of mean-field equation
of $x$ in the standard SIS model (a) and the SIS$m$ ($m\geq2$) model
(b). The solid and dashed lines indicate the stable and unstable
solution of $x$, respectively. \label{fig1}}
\end{figure}

For the standard SIS model, it is well-known that the model
undergoes a continuous phase transition from an epidemic-free state
to an epidemic-spreading state at $\lambda=\lambda_{c(1)}\equiv 1$,
as shown in Fig.\ref{fig1}(a). For $\lambda<\lambda_{c(1)}$, the
only stable steady solution of $x$ is $x_s=0$. For
$\lambda>\lambda_{c(1)}$, the solution $x_s=0$ becomes unstable and
a new stable solution of epidemics emerges
$x_s=1-\lambda_{c(1)}/\lambda$.

\begin{figure}
\centerline{\includegraphics*[width=0.8\columnwidth]{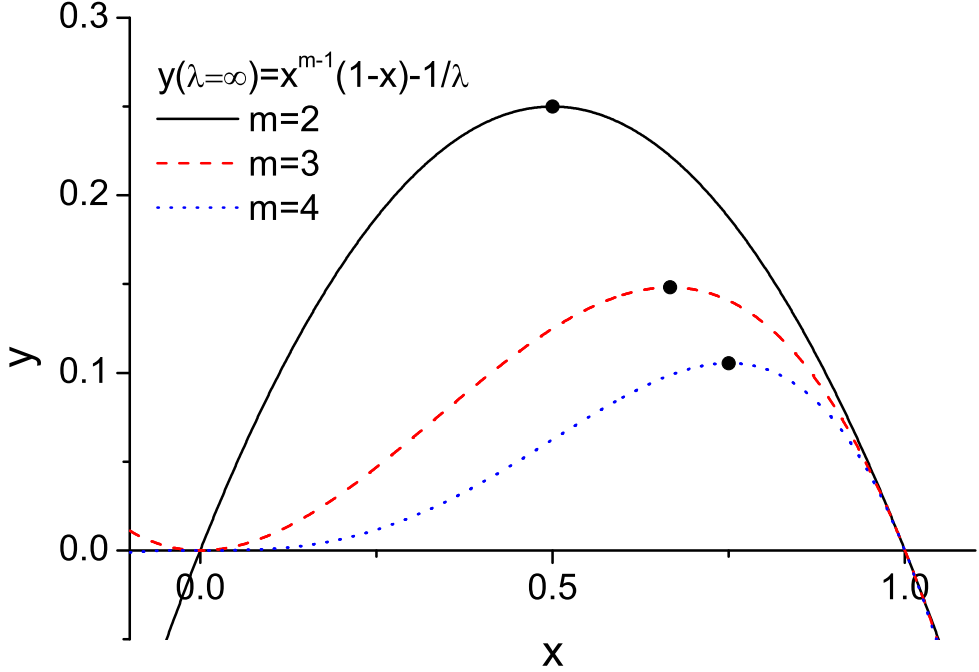}}
\caption{(color online). $y=x^{m-1}(1-x)-1/\lambda$ at
$\lambda=\infty$ as a function of $x$ for several $m$. The solid
circles indicate the locations of the maxima in $y$ for $x\in (0,1)$
\label{fig2}}
\end{figure}

For the SIS$m$ model with $m\geq2$, we find that, by linear
stability analysis, the solution $x_s=0$ is always stable regardless
of the value of $\lambda$, which is significantly different from the
standard SIS model. Moreover, in order to seek nonvanishing
solutions of $x$, it yields, by Eq.\ref{eq1},
$x^{m-1}(1-x)=1/\lambda$, or equivalently, $y=0$ where $y$ is
defined as $y=x^{m-1}(1-x)-1/\lambda$. Firstly, in Fig.\ref{fig2} we
plot $y$ as a function of $x$ at $\lambda=\infty$ ($1/\lambda=0$)
for several distinct values of $m$. Clearly, we find that $y$ has a
maximum $y_m=(m-1)^{m-1}/m^m$ at $x=x_m=(m-1)/m$. We then consider a
finite $\lambda>0$, in which the curves $y\sim x$ will be overall
shifted downward with a quantity $1/\lambda$. If $\lambda$ is large
enough, the shift is so small that there exist two solutions of $x$
between 0 and 1, denoted by $x_s^\pm$ ($x_s^+>x_s^-$). By linear
stability analysis, one can easily check that $x_s^+$ is stable and
$x_s^-$ is unstable. If $\lambda$ is decreased from above, the shift
will become large enough until the two solutions vanish. At the
critical case, $y_m=1/\lambda$, leading to the critical value of
$\lambda$ for $m\geq2$,
\begin{eqnarray}
{\lambda _{c(m)}} = \frac{{{m^m}}}{{{{(m - 1)}^{m - 1}}}}{\kern 1pt}
{\kern 1pt} {\kern 1pt} {\kern 1pt} {\kern 1pt} {\kern 1pt} {\kern
1pt} {\kern 1pt} {\kern 1pt} {\kern 1pt} {\kern 1pt} {\kern 1pt}
{\kern 1pt} ({\kern 1pt} {\kern 1pt} m \geqslant 2) \label{eq2}
\end{eqnarray}
The bifurcation diagrams for $m\geq 2$ are summarized in
Fig.\ref{fig1}(b). For $\lambda>\lambda_{c(m)}$, there are two
stable solutions, $x_s=0$ and $x_s=x_s^+$, and an unstable solution,
$x_s=x_s^-$. Depending on the initial density of infected
individuals, the model will arrive at either an epidemic-free state
($x_s=0$) or an epidemic-spreading state ($x_s=x_s^+$). As $\lambda$
decreases, $x_s^+$ and $x_s^-$ get close to each other, until they
colloid and annihilate at $\lambda=\lambda_{c(m)}$ via a saddle-node
bifurcation. For $\lambda<\lambda_{c(m)}$, the only stable solution
is $x_s=0$. Concretely, for the SIS$2$ model,
\begin{eqnarray}
x_s^ \pm  = \frac{1}{2} \pm \frac{1}{2}\sqrt {\frac{{\lambda  -
{\kern 1pt} {\lambda _{c(2)}}}}{\lambda },} {\kern 1pt} {\kern 1pt}
{\kern 1pt} {\kern 1pt} {\kern 1pt} {\kern 1pt} {\kern 1pt} {\kern
1pt} {\kern 1pt} {\kern 1pt} {\kern 1pt} {\kern 1pt} {\kern 1pt}
{\kern 1pt} and{\kern 1pt} {\kern 1pt} {\kern 1pt} {\kern 1pt}
{\kern 1pt} {\kern 1pt} {\kern 1pt} {\kern 1pt} {\kern 1pt} {\kern
1pt} {\kern 1pt} {\kern 1pt} {\kern 1pt} {\kern 1pt} {\kern 1pt}
{\kern 1pt} {\kern 1pt} {\kern 1pt} {\kern 1pt} {\lambda _{c(2)}} =
4 \label{eq3}
\end{eqnarray}
For the SIS$3$ model,
\begin{eqnarray}
x_s^ \pm  = \frac{1}{3} + \frac{1}{3}\cos \frac{\theta }{3} \pm
\frac{{\sqrt 3 }}{3}\sin {\kern 1pt} \frac{\theta }{3},{\kern 1pt}
{\kern 1pt} {\kern 1pt} {\kern 1pt} {\kern 1pt} {\kern 1pt} {\kern
1pt} {\kern 1pt} {\kern 1pt} {\kern 1pt} {\kern 1pt} {\kern 1pt}
{\kern 1pt} {\kern 1pt} and{\kern 1pt} {\kern 1pt} {\kern 1pt}
{\kern 1pt} {\kern 1pt} {\kern 1pt} {\kern 1pt} {\kern 1pt} {\kern
1pt} {\kern 1pt} {\kern 1pt} {\kern 1pt} {\kern 1pt} {\kern 1pt}
{\kern 1pt} {\kern 1pt} {\kern 1pt} {\kern 1pt} {\kern 1pt} {\kern
1pt} {\lambda _{c(3)}} = \frac{{27}}{4} \label{eq4}
\end{eqnarray}
where $\theta  = \arccos \left( {{{2{\lambda _{c(3)}}}
\mathord{\left/ {\vphantom {{2{\lambda _{c(3)}}} \lambda }} \right.
 \kern-\nulldelimiterspace} \lambda } - 1} \right)$.

\section{Master equation and WKB approximation}
Since the mean-field treatment ignores the effect of stochastic fluctuations, it fails to account for the process of
epidemic extinction induced by large fluctuations. To the end, let us define by $P_n(t)$ the probability that the number of
infected individuals is $n$ at time $t$. The master equation for
$P_n(t)$ reads,
\begin{eqnarray}
\frac{{\partial {P_n}\left( t \right)}}{{\partial t}} = {W_ + }(n -
1){P_{n - 1}}\left( t \right) + {W_ - }(n + 1){P_{n + 1}}\left( t
\right) - \left[ {{W_ + }(n) + {W_ - }(n)} \right]{P_n}\left( t
\right) \label{eq5}
\end{eqnarray}
where $W_ + (n)$ ($W_ - (n)$) is the infection (recovery) rate, given by
\begin{eqnarray}
{W_ + }(n) = \lambda (N - n)\prod\limits_{l = 0}^{m - 1} {\frac{{n -
l}}{N}} \label{eq6}
\end{eqnarray}
and
\begin{eqnarray}
{W_ - }(n) = n \label{eq7}
\end{eqnarray}

By employing the WKB approximation for the probability
\cite{PhysRevLett.101.078101,JSP127.861}, we write
\begin{eqnarray}
P(n) = {e^{ - N\mathcal {S}(x)}}. \label{eq8}
\end{eqnarray}
As usual, we assume $N$ is large and take the leading order in a
$N^{-1}$ expansion, by writing $P(n \pm 1) \approx P(n){e^{ \mp
{{\partial \mathcal {S}} \mathord{\left/
 {\vphantom {{\partial \mathcal{S}} {\partial x}}} \right.
 \kern-\nulldelimiterspace} {\partial x}}}}$ and $W(n \pm 1) \approx W(n)$.
We can arrive at the Hamilton-Jacobi equation
\cite{PhysRevLett.101.078101,JSP127.861},
\begin{eqnarray}
\frac{{\partial \mathcal {S}}}{{\partial t}} + \mathcal {H}(x,p) = 0
\label{eq9}
\end{eqnarray}
where $\mathcal {S}$ and $\mathcal {H}$ are called the action and
Hamiltonian, respectively. As in classical mechanics, the
Hamiltonian is a function of the coordinate $x$ and its conjugate
momentum $p = {{\partial \mathcal {S}} \mathord{\left/
 {\vphantom {{\partial \mathcal {S}} {\partial x}}} \right.
 \kern-\nulldelimiterspace} {\partial x}}$,
\begin{eqnarray}
\mathcal {H}(x,p) = {w_ + }(x)\left( {{e^p} - 1} \right) + {w_ -
}(x)\left( {{e^{ - p}} - 1} \right) \label{eq10}
\end{eqnarray}
where ${w_ \pm }(x) = {{{W_ \pm }(n)} \mathord{\left/
 {\vphantom {{{W_ \pm }(n)} N}} \right.
 \kern-\nulldelimiterspace} N}$ are the reaction rates per person. Substituting Eq.\ref{eq6} and Eq.\ref{eq7} to Eq.\ref{eq10}, we obtain
\begin{eqnarray}
\mathcal {H}(x,p) = \lambda {x^m}\left( {1 - x} \right)\left( {{e^p}
- 1} \right) + x\left( {{e^{ - p}} - 1} \right) \label{eq11}
\end{eqnarray}
We then write the canonical equations of motion,
\begin{eqnarray}
\dot x = {\partial _p}\mathcal {H}(x,p) = \lambda {x^m}\left( {1 -
x} \right){e^p} - x{e^{ - p}}{\kern 1pt} \label{eq12}
\end{eqnarray}
\begin{eqnarray}
\dot p =  - {\partial _x}\mathcal {H}(x,p) =  - \lambda \left[
{m{x^{m - 1}} - (m + 1){x^m}} \right]\left( {{e^p} - 1} \right) -
{e^{ - p}} + 1 \label{eq13}
\end{eqnarray}
We are interested in the extinction trajectory from an epidemic
state to an extinct state of epidemics. This implies that there will
be some trajectory along which $\mathcal {S}$ is minimized, which
represents the maximal probability of such an extinction event. This
corresponds to the zero-energy ($\mathcal {H}=0$) trajectory in the
phase space $(x, p)$ from an epidemic fixed point $A$ to an
extinction one $C$. According to Eq.\ref{eq11}, the condition
$\mathcal {H}=0$ implies that there are three lines: $x=0$, $p=0$,
and
\begin{eqnarray}
\lambda {x^m} - \lambda {x^{m - 1}} + e^{-p} = 0. \label{eq14}
\end{eqnarray}
These three lines determine the topology of the optimal extinction
path on the phase plane. Especially, the line $p=0$ corresponds to
the result of mean-field treatment, as Eq.\ref{eq12} for $p=0$
recovers to the mean-field equation \ref{eq1}.

\begin{figure}
\centerline{\includegraphics*[width=1.0\columnwidth]{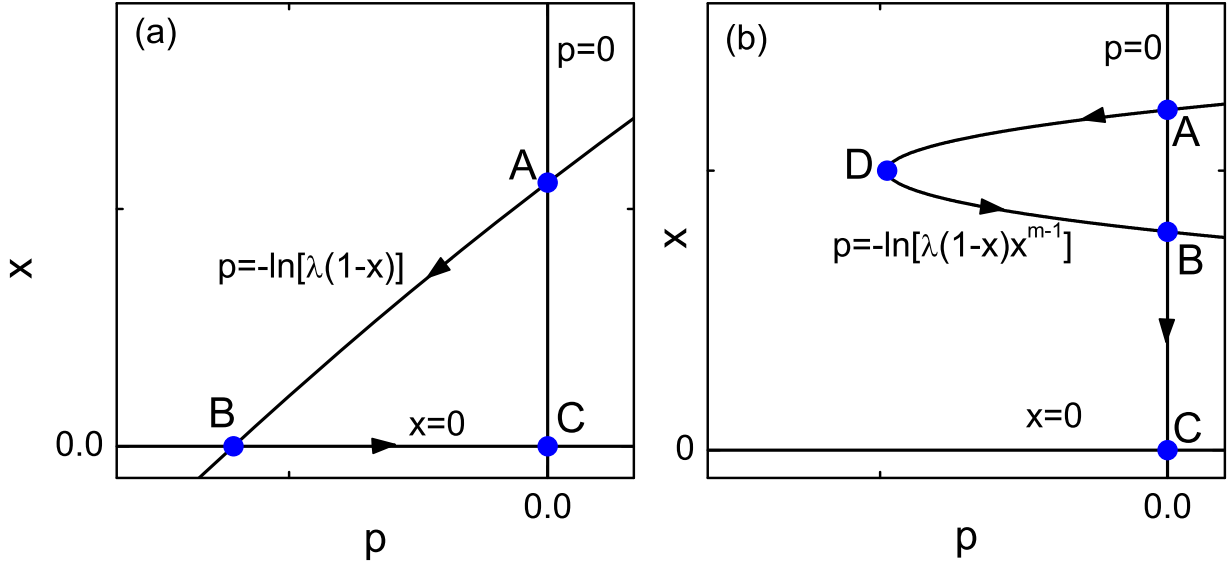}}
\caption{(color online). The optimal path of extinction on the $(x,
p)$ plane for the standard SIS model (a) and the SIS$m$ ($m\geq2$)
model (b). \label{fig3}}
\end{figure}

Fig.\ref{fig3}(a) depicts the most probable trajectory of extinction
in the standard SIS model from $A$ to $C$ via a fluctuational fixed
point $B$. The coordinates on the phase space $(x, p)$ of these
points are: $A = (1 -\lambda_{c(1)}/\lambda,0)$, $B=(0, -\ln
\lambda)$, and $C=(0, 0)$. In general, the average time to reach
extinction $\left\langle T \right\rangle$ is inversely proportional
to the probability of the event, and thus can be related to our
action by $\left\langle T \right\rangle  \sim {e^{N\mathcal {S}}}$,
where $\mathcal {S}$ is given by
\begin{eqnarray}
\mathcal {S}(m=1) = \int {pdx}  =  \ln \lambda  + \frac{1}{\lambda }
- 1 \hfill \label{eq15}
\end{eqnarray}

In Fig.\ref{fig3}(b), we show the typical results for $m\geq 2$. The
points $A$, $B$, and $C$ locate at the mean-field line, and their
coordinates $(x, p)$ are:  $A = ({x_s^+},0)$, $B=(x_s^-, 0)$, and
$C=(0, 0)$. The action $S$ can be analytically determined by

\begin{eqnarray}
\mathcal {S}(m\geq2) = \int_{x_s^ - }^{x_s^ + } {\ln \left[ {\lambda
{x^{m - 1}}(1 - x)} \right]dx}  = F(x_s^ + )-F(x_s^ -) \label{eq16}
\end{eqnarray}
with
\begin{eqnarray}
F(x)= {(\ln \lambda  - m)x + (m - 1)x \ln x - (1 - x) \ln(1 - x)}
\label{eq17}
\end{eqnarray}


\section{Scaling in weak infection regime}
In the following, we will derive the scaling of the action $\mathcal
{S}$ with respect to the distance to the critical infection rate
$\eta=\lambda-\lambda_c$ in the weak infection regime (i.e., $\eta
\sim \mathcal{O}(1)$). For the standard SIS model close to
$\lambda_{c(1)}$, the paths to extinction are approximately linear
from $A$ to $B$, and thus the action $\mathcal {S}$ can be
approximately calculated by
\cite{PhysRevLett.101.078101,JSM2009P01005}
\begin{eqnarray}
\mathcal {S} (m=1) \approx \frac{1}{2}\frac{{\lambda  - {\lambda
_{c(1)}}}}{\lambda }\ln \lambda  \approx \frac{1}{2}{\left(
{\frac{{\lambda  - {\lambda _{c(1)}}}}{\lambda }} \right)^2} \sim
{\eta ^2}  \label{eq18}
\end{eqnarray}

For $m\geq2$ and the infection rate close to $\lambda_{c(m)}$, the
paths to extinction from $A$ to $B$ are approximately regarded as
two lines connected by a point $D$, where $D=(x_m,
-\ln(\lambda/\lambda_{c(m)}))$ corresponds to the peak of the curve
described by Eq.\ref{eq14}. Therefore, the action $\mathcal {S}$
approximately equals to $\mathcal {S} \approx \frac{1}{2}(x_s^ +  -
x_s^ - )\ln(\lambda/\lambda_{c(m)})$, where $x_s^+$ and $x_s^-$ are
the solutions of the equation $x^{m-1}(1-x)=1/\lambda$. Since
$\lambda$ is close to $\lambda_{c(m)}$, the two solutions are very
close to $x_m$, and thus we make an expansion
$x_s^{\pm}=x_m\pm\epsilon$. Inserting the expansion to the above
equation, we obtain $x_s^ +  - x_s^ - \approx 2\sqrt {{{(\lambda  -
{\lambda _{c(m)}})} \mathord{\left/
 {\vphantom {{(\lambda  - {\lambda _{c(m)}})} {\lambda {\lambda _{c(m)}}}}} \right.
 \kern-\nulldelimiterspace} {\lambda {\lambda _{c(m)}}}}}$. Thus, we arrive at the scaling relation for $m\geq2$,
\begin{eqnarray}
\mathcal {S}(m\geq2) \approx \sqrt {\frac{{\lambda  - {\lambda
_{c(m)}}}}{{\lambda {\lambda _{c(m)}}}}} \frac{{\lambda  - {\lambda
_{c(m)}}}}{{ {\lambda _{c(m)}}}} \sim {\eta ^{\frac{3}{2}}}
\label{eq19}
\end{eqnarray}
This is consistent with the standard scaling of the activation
energy of escape near a saddle-node bifurcation point
\cite{JCP100.5735}.

\section{Numerical validation}
In order to validate the theoretical results, we have made the
stochastic simulation for the master equation (5) by Gillespie's
algorithm \cite{JCP22.403,JPC81.2340}. The main idea of Gillespie's
algorithm is to randomly determine what the succeeding reaction step
is and when it will happen according to the transition rates
$W_{\pm}(n)$. However, epidemic extinction is a rare event that
occurs very infrequently, especially for large $\lambda$ or $N$.
Thus, the conventional brute-force simulation becomes prohibitively
inefficient. To overcome this difficulty, we have employed a
recently developed rare-event sampling method, forward flux sampling
(FFS) \cite{PRL05018104,JPH09463102}, combined with Gillespie's
algorithm. The FFS method first defines an order parameter to
distinguish between the initial epidemic-spreading state
$\mathcal{I}$ and the final epidemic-free state $\mathcal{F}$, and
then uses a series of interfaces to force the system from
$\mathcal{I}$ to $\mathcal{F}$ in a ratchet-like manner. Here, it is
convenient to select the number of infected individuals $n$ as the
order parameter. We define that the system is in $\mathcal{I}$ if
$n=n_0$ and it is in $\mathcal{F}$ if $n=0$, where $n_0=N x_s$ and
$x_s$ is the steady density of infected individuals at $\mathcal{I}$
 state. A series of nonintersecting interfaces $n_i$ ($0<i<N_{in}$) lie
between states $\mathcal{I}$ and $\mathcal{F}$, such that any path
from $\mathcal{I}$ to $\mathcal{F}$ must cross each interface
without reaching $n_{i+1}$ before $n_i$, where $N_{in}$ is the
number of the interfaces. The algorithm first runs a long-time
simulation which gives an estimate of the flux
$\Phi_{\mathcal{I},0}$ escaping from the basin of $\mathcal{I}$ and
generates a collection of configurations corresponding to crossings
of the interface $n_0$. The next step is to choose a configuration
from this collection at random and use it to initiate a trial run
which is continued until it either reaches $n_1$ or returns to
$n_0$. If $n_1$ is reached, store the configuration of the end point
of the trial run. Repeat this step, each time choosing a random
starting configuration from the collection at $n_0$. The fraction of
successful trial runs gives an estimate of of the probability of
reaching $n_1$ without going back into $\mathcal{I}$, $P\left( {n_1
|n_0} \right)$. This process is repeated, step by step, until
$n_{N_{in}}$ is reached, giving the probabilities $P\left( {n_{i+1}
|n_i} \right)$ ($i=1, \cdots, N_{in}-1$). Finally, we get the mean
extinction time,
\begin{equation}
\left\langle T \right\rangle  = \frac{1}{\Phi_{\mathcal{I},0}
\prod\nolimits_{i=0}^{N_{in}-1}{P\left( {n_{i + 1} |n_i} \right)}}
\end{equation}

\begin{figure}
\centerline{\includegraphics*[width=1.0\columnwidth]{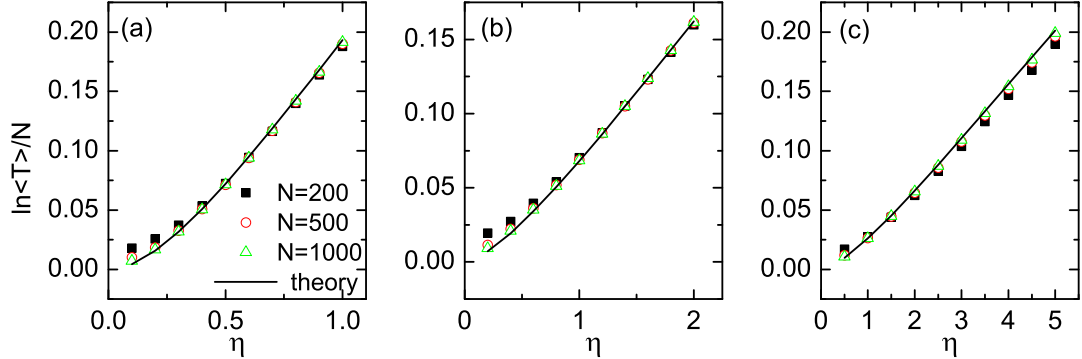}}
\caption{(color online). $\ln\langle T \rangle/N$ as a function of
$\eta=\lambda-\lambda_c(m)$ for $m=1$ (a), $m=2$ (b), and $m=3$ (c).
The lines indicate the theoretical results of the action $\mathcal
{S}$ from Eq.\ref{eq15} and Eq.\ref{eq16}. \label{fig4}}
\end{figure}

In Fig.\ref{fig4}(a-c), we show $\ln\langle T \rangle/N$ as a
function of $\eta=\lambda-\lambda_c(m)$ for $m=1$, 2, and 3,
respectively. Since $\langle T \rangle  \sim {e^{N\mathcal {S}}}$,
the action $\mathcal {S}$ is comparable to $\ln\langle T \rangle/N$.
For comparison, we plot the theoretical results of $\mathcal {S}$
(Eq.\ref{eq15} and Eq.\ref{eq16}), indicated by the lines in
Fig.\ref{fig4}. For large $N$, there are excellent agreements
between the simulations and theoretical predictions for all
$\lambda$. For small $N$, the agreements hold only for large
$\lambda$. While for small $N$ and $\lambda$, the theory disagrees
the simulations. This is because that in this case $\langle T
\rangle$ is not very long, such that $e^{N\mathcal {S}}$ can be
comparable to its pre-exponential factor which was not considered in
our analysis \cite{PhysRevE.81.021116}.

\begin{figure}
\centerline{\includegraphics*[width=1.0\columnwidth]{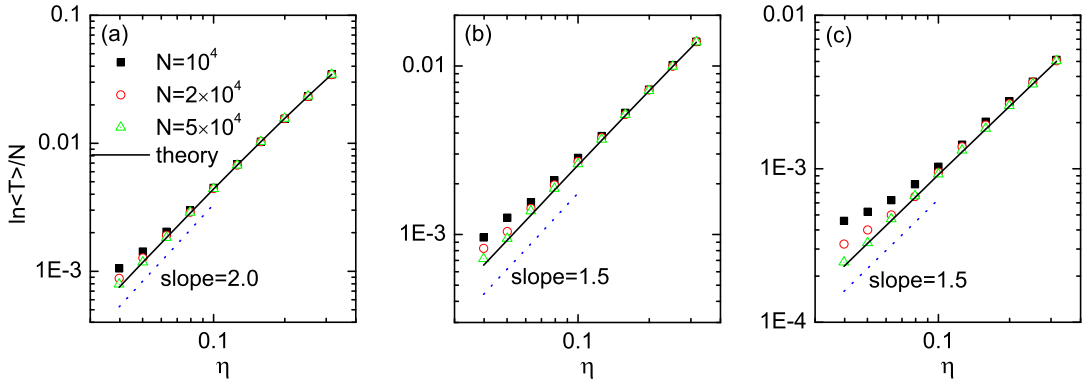}}
\caption{(color online). Double logarithmic plot of $\ln\langle T
\rangle/N$ as a function of $\eta=\lambda-\lambda_c(m)$ for $m=1$
(a), $m=2$ (b), and $m=3$ (c). The solid lines indicate the
theoretical results of the action $\mathcal {S}$ from Eq.\ref{eq15}
and  Eq.\ref{eq16}. The dotted lines indicate the reference slopes
of the scaling near the epidemic thresholds from Eq.\ref{eq18} and
Eq.\ref{eq19}. \label{fig5}}
\end{figure}

In the following, we shall show the numerical verification of the
scaling relation near the bifurcation point. Along with
Eq.\ref{eq18} and Eq.\ref{eq19}, $\ln\langle T \rangle/N$ is
expected to exhibit the scaling to $\eta$ with an exponent $2$ for
$m=1$ and $3/2$ for $m\geq2$ near the epidemic threshold.
Fig.\ref{fig5} shows $\ln\langle T \rangle/N$ as a function of
$\eta$ on double logarithmic coordinates for $m=1$, 2, and 3,
respectively, in which we have used quite large $N$ to reduce the
contribution of the pre-exponential factor to $\langle T \rangle$.
The simulation results agree well with the theoretical predictions
from Eq.\ref{eq18} and Eq.\ref{eq19}.

\section{Conclusions and Discussion}
In conclusion, we have employed the WKB method to study the epidemic
extinction process in a generalized SIS model (SIS$m$). The method
transfers the master equation to an effective Hamilton system. This
allows us to analytically determine the mean extinction time of
epidemics, which is exponentially dependent on the population size
$N$ and the action $\mathcal {S}$ along the zero-energy trajectories
from the epidemic point to the extinction point on the phase space.
Since the properties of bifurcation for $m=1$ and $m\geq2$ are
different, we independently derive the analytical form of $\mathcal
{S}$ for the two cases. Near the epidemic bifurcation point, we
obtain the scaling of $\mathcal {S}$ with the distance to the
bifurcation $\eta$, $\mathcal {S} \sim \eta^2 $ for $m=1$ and
$\mathcal {S} \sim \eta^{3/2} $ for $m\geq2$. Finally, we have used
a rare-event simulation method, FFS, to validate the theoretical
results. While the model under study is a natural extension of the
standard SIS model, a canonical model of epidemic spreading, it may
be relevant to synergistic effects on the dynamics of opinion
formation \cite{Science329.1194,PhysRevLett.106.218701}. If a group
of neighboring persons to a selected person unanimously shares an
opinion, the group pressure causes the selected person to take the
opinion of the group. Therefore, the present theoretical study could
provide useful understanding for the large fluctuation effect on the
opinion dynamics. In the future, it would be desirable to consider
the effect of interacting patterns among agents on the extinction of
the generalized SIS model
\cite{EPL108.58008,PhysRevLett.117.028302}.

\begin{acknowledgments}
This work was supported by National Science Foundation of China
(Grants Nos. 11205002, 61473001, 11405001), the Key Scientific
Research Fund of Anhui Provincial Education Department (Grant No.
KJ2016A015) and ``211" Project of Anhui University (Grant No.
J01005106).
\end{acknowledgments}

%

\end{document}